\documentclass[journal,10pt]{IEEEtran}

\usepackage{cite}

\usepackage{graphicx} 
\usepackage{multirow}
\usepackage{adjustbox}
\usepackage{multicol}
\usepackage{lipsum}
\usepackage[numbers,square, comma, sort & compress]{natbib}
\usepackage[hidelinks]{hyperref}
\usepackage{xcolor,soul,lipsum}
\newcommand{\myul}[2][black]{\setulcolor{#1}\ul{#2}\setulcolor{black}}
\usepackage{amssymb,amsmath,enumerate}
\usepackage{dblfloatfix}
\usepackage{array}
\newcolumntype{C}[1]{>{\centering\let\newline\\\arraybackslash\hspace{0pt}}m{#1}}
\ifCLASSOPTIONcompsoc
\usepackage[caption=false,font=footnotesize,labelfont=sf,textfont=sf]{subfig}
\else
\usepackage[caption=false,font=footnotesize]{subfig}
\fi

\renewcommand{\baselinestretch}{0.95}
\newcolumntype{C}[1]{>{\centering\let\newline\\\arraybackslash\hspace{0pt}}m{#1}}

\newcommand{\Zstroke}{%
  \text{\ooalign{\hidewidth\raisebox{0.2ex}{--}\hidewidth\cr$Z$\cr}}%
}
\newcommand{\zstroke}{%
  \text{\ooalign{\hidewidth -\kern-.3em-\hidewidth\cr$z$\cr}}%
}

\begin{document}
 \title{Design and Implementation of Low Complexity Reconfigurable Filtered-OFDM based LDACS}
 \author{Niharika Agrawal, Abhishek Ambede, ~S.~J.~Darak, ~A.~P.~Vinod, and ~A.~S.~Madhukumar
 	\thanks{Niharika Agrawal and Abhishek Ambede are joint first-authors.}
 	\thanks{Niharika Agrawal and S.~J.~Darak are with ECE Department, IIIT-Delhi, India 110020, e-mail: \{niharikaa, sumit\}@iiitd.ac.in.}
 	\thanks{Abhishek Ambede and ~A.~S.~Madhukumar are with SCSE, NTU, Singapore 639798, e-mail: abhishek7@e.ntu.edu.sg, asmadhukumar@ntu.edu.sg.}
 	\thanks{A. P. Vinod is with EE Department, IIT-Palakkad, India 678557, e-mail: vinod@iitpkd.ac.in.}
 }
\maketitle

\begin{abstract}
        $L$-band Digital Aeronautical Communication System (LDACS) aims to exploit vacant spectrum in $L$-band via spectrum sharing, and orthogonal frequency division multiplexing (OFDM) is the currently accepted LDACS waveform. Recently, various works dealing with improving the spectrum utilization of LDACS via filtering/windowing are being explored. In this direction, we propose an improved and low complexity reconfigurable filtered OFDM (LRef-OFDM) based LDACS using novel interpolation and masking based multi-stage digital filter. The proposed filter is designed to meet the stringent non-uniform spectral attenuation requirements of LDACS standard. It offers significantly lower complexity as well as higher transmission bandwidth than state-of-the-art approaches. We also integrate the proposed filter in our end-to-end LDACS testbed realized using Zynq System on Chip and analyze the performance in the presence of $L$-band legacy user interference as well as LDACS specific wireless channels. Via extensive experimental results, we demonstrate the superiority of the proposed LRef-OFDM over OFDM and Filtered-OFDM based LDACS in terms of power spectral density, bit error rate, implementation complexity, and group delay parameters. 
		
	
 \end{abstract}


%
\IEEEpeerreviewmaketitle
\vspace{-0.17cm}
\section{Introduction}
The global air traffic has increased tremendously over the past few decades and numerous studies and statistical forecasts predict its rapid growth in the future as well \cite{1,2}. The consequent massive growth in the air-ground communication traffic has led to congestion in currently used very high frequency (VHF) aeronautical communication band (118-137 MHz). To ensure reliable communications with the desired quality-of-service and to support the growing need of multimedia communications that demand high bandwidth (BW),  a new L-band (960-1164 MHz) Digital Aeronautical Communication System (LDACS) \cite{3} is under development. In this paper, we focus on the physical layer (PHY) of LDACS.

The aeronautical $L$-band is already occupied by multiple legacy systems, most prominently by the distance measuring equipment (DME), which operates in the range 962–1213 MHz. LDACS is proposed to exploit the 1 MHz vacant band(s) between adjacent DME users via inlay based spectrum sharing approach. Orthogonal frequency division multiplexing (OFDM), widely used in cellular and Wi-Fi standards, is the preferred choice for LDACS PHY waveform. However, LDACS environment has always-ON legacy DME signals, and high out-of-band attenuation (OOBA) of OFDM limits the transmission BW to only 498 kHz out of the available 1 MHz spectrum (50\% spectral utilization) \cite{3}. In this direction, we proposed filtered-OFDM (FOFDM) based LDACS to improve the OOBA, thereby allowing higher transmission BWs of up to 732 kHz without compromising on bit-error-rate (BER), and our frame structure is compatible with legacy LDACS \cite{arxn, Nr}.

In this paper, we focus on the complexity aspect of FOFDM based LDACS transceivers especially for deployment onboard aircrafts, along with in-depth performance analysis of end-to-end LDACS transceiver on heterogeneous Zynq System on Chip (ZSoC) platform consisting of FPGA and ARM processor. As onboard aircraft systems are battery-powered, reducing the LDACS PHY complexity is an essential step towards extending the battery life. Specifically, we propose a novel interpolation and masking based multi-stage digital finite impulse response (FIR) filter that when integrated with LDACS transceiver, not only meets the stringent non-uniform spectral attenuation requirements of LDACS standard but also allows variable transmission BWs up to 732 kHz. Compared to our previous works, we offer significant savings in the area (12.78 \% in DSP48 units) and power consumption (14.14 \%) at the transceiver level, along with more than 50\% fewer multipliers over conventional single-stage filtering approaches. Hereafter, the proposed solution is referred to as low complexity reconfigurable filtered OFDM (LRef-OFDM) based LDACS. Various LDACS transceivers are implemented on the ZSoC platform and integrated with analog front-end (AFE) for experimental evaluation in the presence of LDACS specific wireless signals and DME interference. Via extensive experimental results, we demonstrate the superiority of the LRef-OFDM over OFDM and Filtered-OFDM (FOFDM) based LDACS in terms of power spectral density (PSD), BER, implementation complexity and group delay parameters for a wide range of signal-to-noise ratios (SNRs) and word-lengths (WLs). We begin with the design of the proposed filter in the next section.

\begin{figure*}[!h]
    \centering
    \captionsetup{justification=centering}
    \includegraphics[width=0.9\linewidth]{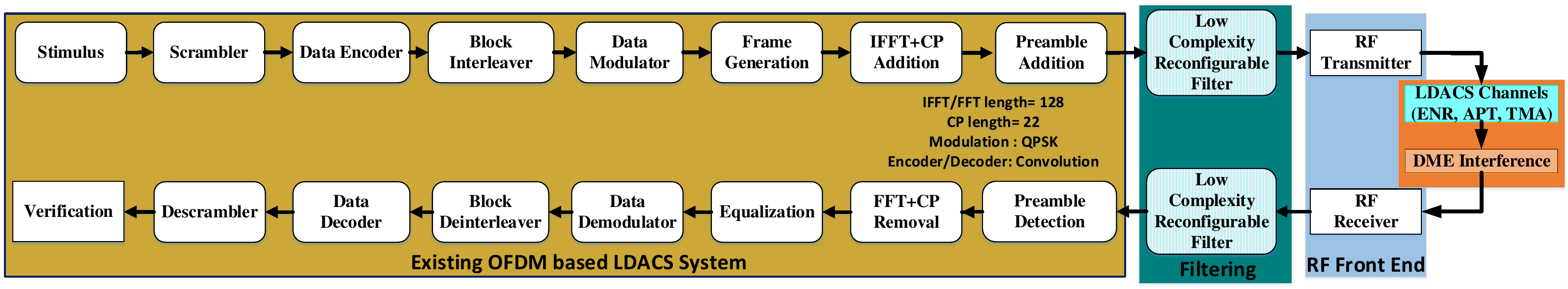}
    \vspace{-0.35cm}
    \caption{Building blocks of the proposed LRef-OFDM based LDACS transceiver along with end-to-end testbed.}
    
    \label{blck}
    \vspace{-0.3cm}
    \end{figure*}
    \vspace{-0.17cm}
\section{Proposed Filter Design}
An end-to-end LRef-OFDM based LDACS transceiver is shown in Fig.~\ref{blck}, and it is based on the standardized LDACS transceiver \cite{3,Nr} in terms of frame, protocol, and building blocks. In this section, we focus on the design of the low complexity BW-reconfigurable digital FIR filter in Fig.~\ref{blck} and the aim is to meet the OOBA specifications of LDACS standard for a wide range of transmission BWs. Please refer to supplementary material in \cite{arxn} and Section III herein for design details of other baseband blocks of the transceiver and the ZSoC testbed, respectively. 

\begin{figure}[!t]
  \vspace{-0.3cm}
    \centering
    \captionsetup{justification=centering}
    \includegraphics[scale=0.425]{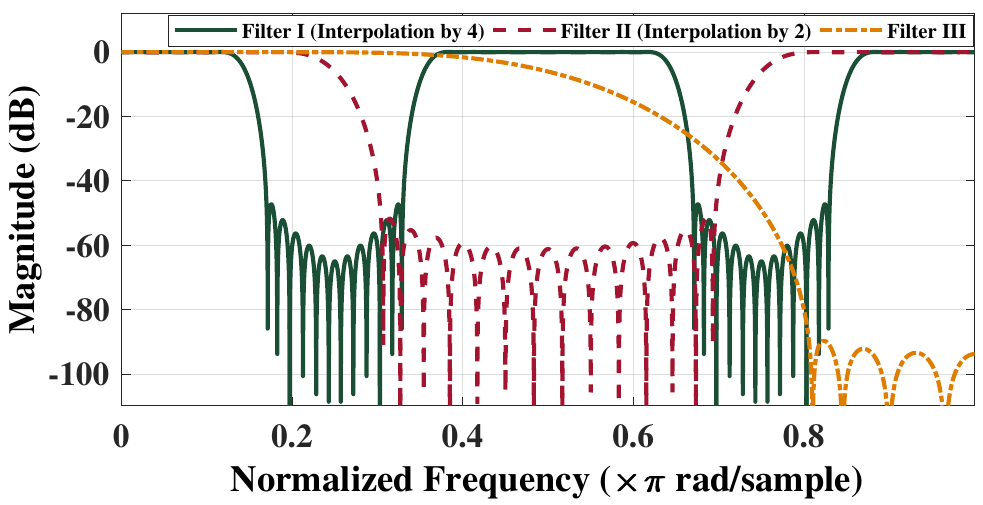}
  
    \vspace{-0.3cm}
    \caption{Frequency responses of sub-filters $H_{I}(z)$, $H_{II}(z)$, $H_{III}(z)$ and the resultant filter $H(z)$ for 498 kHz LDACS transmission BW.}
    \label{resp}
      \vspace{-0.2cm}
\end{figure}

The proposed filter employs the interpolation operation \cite{14,S,W} to reduce the complexity. In this operation, if the coefficients of a lowpass prototype filter are interpolated by a factor $M$, every unit delay in the filter is replaced by $M$ delays. This results in a multi-band frequency response with sub-bands located at even multiples of $\frac{\pi}{M}$, each having its passband and transition BW $\frac{1}{M}$ times that of the prototype filter. The interpolated frequency response can be given by:
\begin{equation}
\vspace{-0.1cm}
    H_{Ip}(z)=\sum_{n=0}^{\frac{N}{2}-1}h_n[z^{-Mn}+z^{-M(N-n)}]+h_{\frac{N}{2}}z^{\frac{-Mn}{2}} 
    \label{1}
    \vspace{-0.1cm}
\end{equation}
where, $h_{0}, h_{1}$ \ldots $h_{\frac{N}{2}}$ are the unique filter coefficients of an $N^{th}$ order FIR filter.
In proposed filter design approach \cite{17}, three sub-filters denoted as Filter I, Filter II, and Filter III are cascaded 
and illustrative frequency responses are shown in Fig.~\ref{resp}. If $H_{I}(z)$, $H_{II}(z)$, $H_{III}(z)$ denote the z-domain representations of sub-filters, the resultant filter is given as,
\begin{equation}
    H(z) = H_{I}(z).H_{II}(z).H_{III}(z)
    \label{2}
\end{equation}

The LDACS signal is over-sampled by a factor of four to assist in interference reduction \cite{3}. Therefore, as the standardized LDACS BW is approximately 500 kHz, the sampling frequency for our filter is chosen as 4 MHz. Fig.~\ref{resp} shows the frequency responses of three sub-filters designed for a transmission bandwidth of 498 kHz on the frequency scale normalized with respect to Nyquist frequency, i.e., half of sampling frequency. The sub-filter design is explained below:

\begin{enumerate}
    \item \textit{Filter I:} The stage I filter is designed with minimum order $(N=26)$ satisfying the stringent spectral mask, and has passband edge ($Fp_{1}$) as $M (=4)$ times of the passband edge of the LDACS signal ($Fp_{s}$) based on its transmission BW, i.e, $Fp_{1}=4*Fp_{s}$. Similarly the stopband edge ($Fs_{1}$) is $4*Fs_{s}$, where $Fs_{s}$ is the stopband edge of the LDACS signal. As the most relaxed required attenuation level is adjacent to the passband, this sub-filter is designed with the most relaxed attenuation specification. The filter response for this stage I filter $H_{I}(z)$ is obtained by substituting the value of I and N in \eqref{1},
    \begin{equation}
    \vspace{-0.1cm}
    H_{I}(z)=\sum_{n=0}^{12}h_n[z^{-4n}+z^{-4(26-n)}]+h_{13}z^{-2n} 
    \label{3}
    \vspace{-0.1cm}
    \end{equation}
    
    \item \textit{Filter II:} This sub-filter removes the unwanted central subband from the frequency response of interpolated Filter I. This filter is designed with order $26$ having interpolation factor $(M)$ of 2, and the filter response $H_{II}(z)$ can be expressed as,
    \begin{equation}
    \vspace{-0.2cm}
    H_{II}(z)=\sum_{n=0}^{12}h_n[z^{-2n}+z^{-2(26-n)}]+h_{13}z^{-n} 
    \label{4}
    \end{equation}
    
    The passband and stopband edge frequencies of Filter II are based on the resultant edge frequencies in the interpolated frequency response of Filter I and can be represented as $Fp_{2}=\frac{F_{m}}{2}$ and $Fs_{2}=1- Fp_{2}$ respectively, where $F_{m}$ is a reference frequency whose value is chosen based on the supported transmission BWs.
    
    \item \textit{Filter III:} This sub-filter has order $14$ and removes the unwanted highpass subband from the cascaded frequency response of Filter I and Filter II, i.e., $H_{I}(z).H_{II}(z)$. The passband and stopband edge frequencies for Filter III  are based on the resultant edge frequencies in the cascaded frequency response and can be represented as $Fp_{3}=\frac{F_{m}}{4}$ and $Fs_{3}=1- Fp_{3}$ respectively. It has the most relaxed transition BW and the most stringent stopband attenuation specification among the three sub-filters. The frequency response of this filter $H_{III}(z)$ can be represented as,
    \begin{equation}
    H_{III}(z)=\sum_{n=0}^{6}h_n[z^{-n}+z^{-(14-n)}]+h_{7}z^{\frac{-n}{2}}
    \label{5}
    \vspace{-0.1cm}
    \end{equation}
    
\end{enumerate}

Here, we mainly focus on the filter design, and the details regarding the complete transceiver are included in an appendix document as supplementary material \cite{arxh}.

\begin{figure}[!h]
     \centering
     \vspace{-0.2cm}
     \subfloat[]{\includegraphics[scale=0.265]{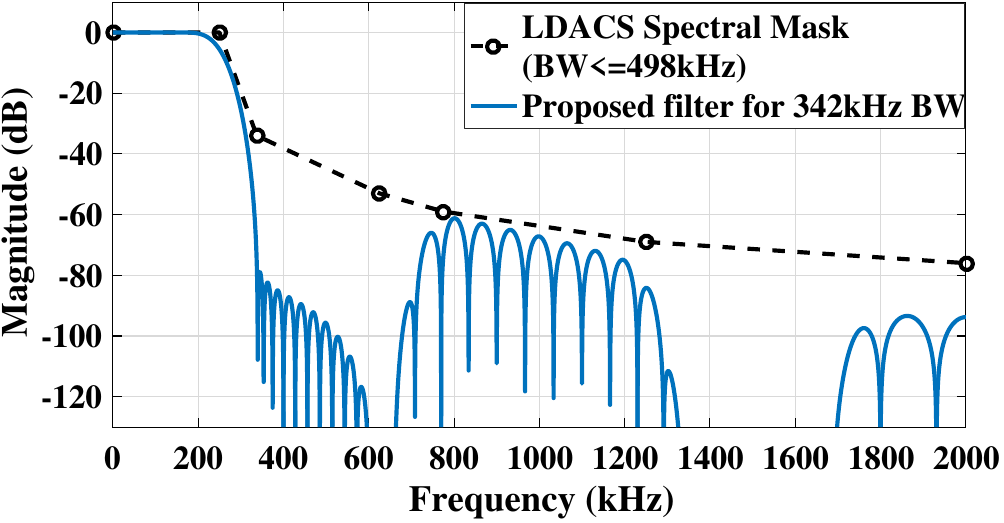}%
         \label{v}}
     \subfloat[]{\includegraphics[scale=0.265]{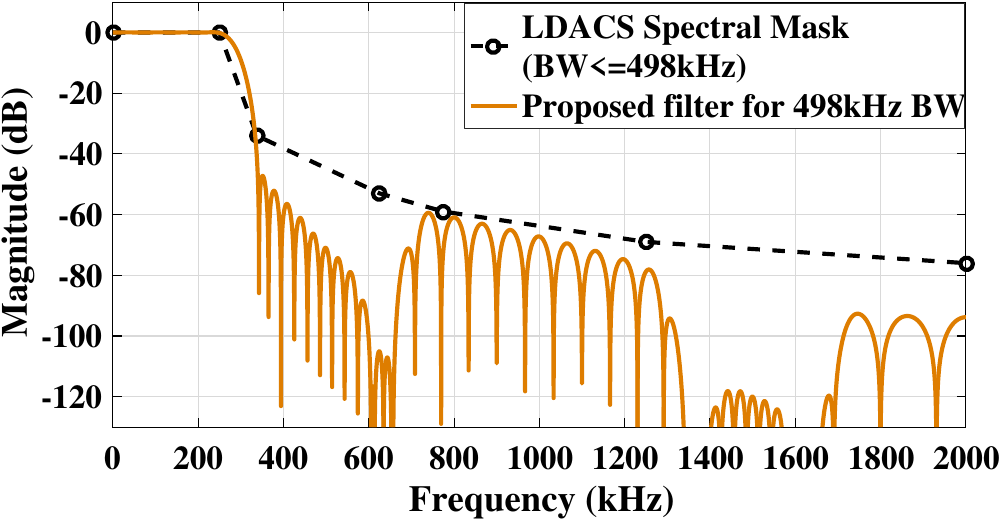}%
         \label{vr4}}
         \vspace{-0.2cm}
         
    \subfloat[]{\includegraphics[scale=0.265]{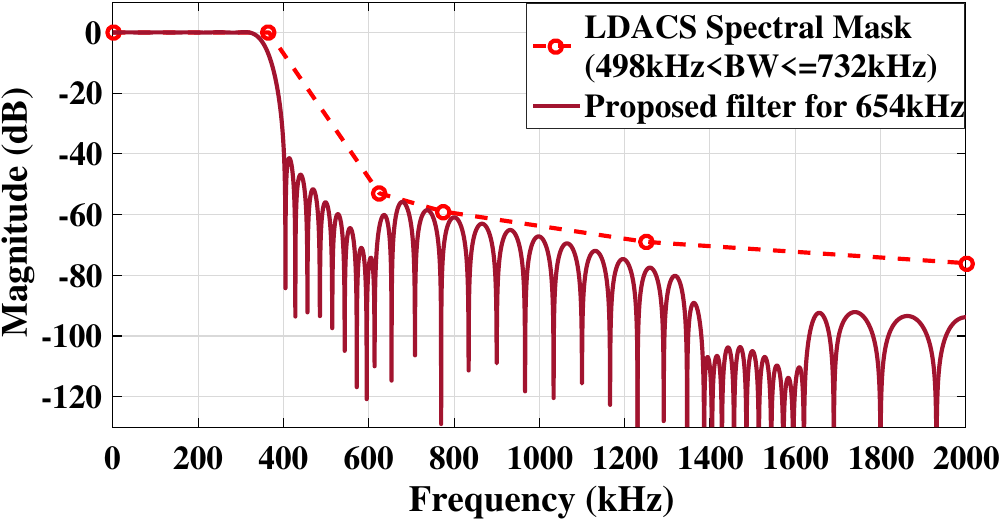}%
         \label{vr4}}
    \subfloat[]{\includegraphics[scale=0.265]{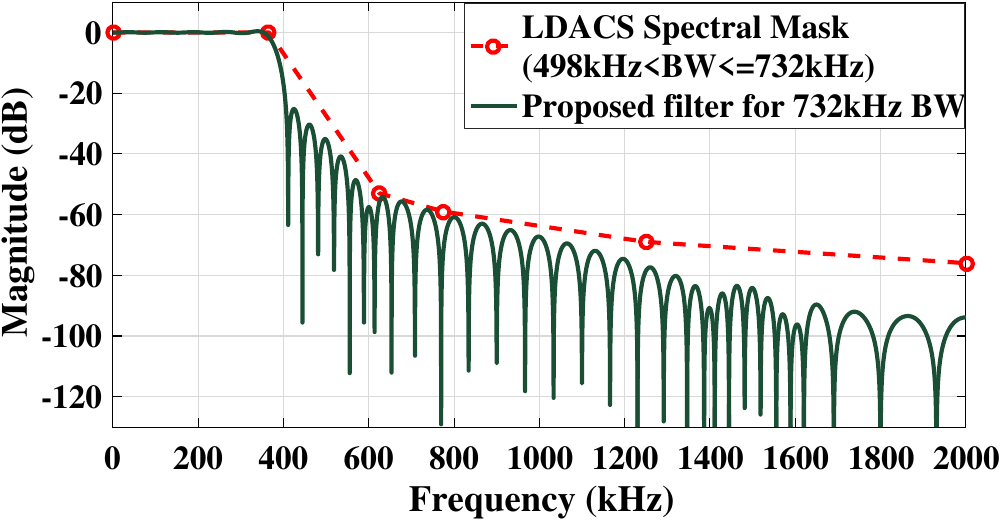}%
         \label{vr4}}
         \vspace{-0.2cm}
     \caption{Overall frequency response characteristics of the proposed filter for the bandwidths: (a) 342 kHz (b) 498 kHz (c) 654 kHz (d) 732 kHz}
     \vspace{-0.25cm}
     \label{fil_r}
 \end{figure}

 In \cite{arxn, Nr}, we showed that LDACS can have additional bandwidths of 342 kHz, 654 kHz, 732 kHz by maintaining compatibility with the standardized frame structure of 498 kHz. To realize a BW reconfigurable transceiver, we store the unique filter coefficients corresponding to all four BWs in memory. We can thus support different transmission BWs on the fly by just selecting appropriate filter coefficients from memory at run-time. This would require a memory component with storage capacity of 144 coefficients: 14*4 (Filter I with order 26) + 14*4 (masking Filter II with order 26) + 8*4 (masking Filter III with order 14). However, to reduce this requirement, we design and use the same masking filters (Filter II and III) for all the BWs, and they are designed as halfband FIR filters to minimize the number of distinct non-zero coefficients. Based on the design of Filter I, a reference frequency $F_{m}$ is computed and the passband and stopband edge frequencies of Filter II and Filter III are chosen based on $F_{m}$. The reference frequency $F_{m}$ is selected based on the supported transmission BWs and is chosen to be the stopband edge frequency of Filter I corresponding to the widest supported transmission BW. As a result, we can use the same masking filters II and III for all BWs and the total required memory storage capacity is just 67 coefficients: 14*4 (Filter I with order 26) + 7 (halfband masking Filter II with order 26) + 4 (halfband masking Filter III with order 14). The overall filter responses for all four transmission BWs satisfying the corresponding LDACS spectral masks are shown in Fig.~\ref{fil_r}. Note that the standardized LDACS spectral mask in \cite{3} is considered for BWs 342 kHz and 498 kHz, and an appropriately modified version is considered for other BWs. 

It is empirically computed that among the different possible combinations of interpolation factors, the lowest overall filter complexity (in terms of total number of unique multipliers) is obtained using the interpolation factors 4, 2, 1 respectively for the three sub-filters \cite{arxh}. The design parameters of the three sub-filters for the different transmission BWs are listed in Table~\ref{spec} (normalized with respect to Nyquist frequency). Except for $Fp_{1}$ and $Fs_{1}$, which are set according to the transmission BW, all other filter design specifications are identical for all the BWs. MATLAB is used to obtain coefficients of the three sub-filters corresponding to the design parameters listed in Table~\ref{spec}. 

\begin{table}[!h]
\centering
\captionsetup{justification=centering}
\caption{Filter design specifications for different LDACS transmission BWs.}
 \vspace{-0.2cm}
\begin{tabular}{|m{1.9cm}|m{0.6cm}|m{0.6cm}|m{0.6cm}|m{0.6cm}|m{0.7cm}|m{0.8cm}|}
\hline
\multirow{2}{*}{\textbf{Parameters}} & \multicolumn{4}{c|}{\textbf{Filter I}}         & \multirow{2}{*}{\textbf{Filter II}} & \multirow{2}{*}{\textbf{Filter III}} \\ \cline{2-5}
                                & \textbf{342 kHz} & \textbf{498 kHz} & \textbf{654 kHz} & \textbf{732 kHz} &                            &                             \\ \hline
Order $(N)$                    & 26      & 26      & 26      & 26      & 26                         & 14                          \\ \hline
Passband Freq.         & 0.3418    & 0.498    & 0.6543   & 0.7324    & 0.3975                       & 0.1988                      \\ \hline
Stopband Freq.         & 0.6724    & 0.6724    & 0.795   & 0.795    & 0.6025                       & 0.8013                      \\ \hline
Attenuation (dB)                & -70.5   & -37.9   & -31.5   & -14.5   & -43.1                        & -81.8                         \\ \hline
 Interpolation $M$            & 4       & 4       & 4       & 4       & 2                          & 1                           \\ \hline
\end{tabular}
\label{spec}
\end{table}
\vspace{-0.2cm}

From Table~\ref{spec}, Fig.~\ref{resp}, and Fig.~\ref{fil_r}, it can be noted that although the attenuation specifications of the different sub-filters are relaxed, the overall cascaded non-uniform frequency responses satisfy the stringent LDACS spectral masks for each transmission BW. This idea of using sub-filters with different relaxed attenuation specifications and thus, lower complexity is a unique feature of this work.

Satisfying the stringent spectral mask specifications \cite{17} should not come at the cost of large delay and high implementation complexity that will also affect the overall power consumption. To understand the filtering complexity, we compare the multiplication complexity (total number of unique multipliers) and group delay of the proposed filter with the FIR filters designed for 498 kHz BW using various state-of-the-art approaches: 1) Parks-McClellan (PM) algorithm \cite{13,arxn}, 2) Least squares (LS) technique \cite{13}, 3) Traditional interpolated FIR (IFIR) technique (wherein two sub-filters are cascaded with only the first subject to interpolation) \cite{14}, and 4) Generalized IFIR technique (wherein more than two sub-filters can be cascaded with multiple out of those subject to interpolation) \cite{15}.

The total number of multipliers involved in the proposed filter is the sum of the multipliers required to implement the three sub-filters. It can be noted that while implementing the FIR sub-filters, the symmetry of their coefficients can be exploited such that only half of the coefficients need to be implemented, using the transposed direct-form FIR filter architecture \cite{19}. Also, for halfband FIR filters (sub-filters II and III in our case), every alternate coefficient is zero, and the central coefficient is always 0.5, which can be implemented simply with a logical shift operation. Exploiting these properties, sub-filters I, II, and III in the proposed LRef-OFDM based LDACS can be implemented using 14, 7, and 4 multipliers respectively and the total number of multipliers required is thus 25. As the three sub-filters are cascaded, the group delay of the proposed filter is the sum of the group delays of sub-filters I, II, and III. The total group delay of the proposed filter is thus (13x4) + (13x2) + (7x1) = 85 samples. In units of time, this corresponds to 21.25 $\mu$s as the sampling frequency is 4 MHz. The total number of required multipliers and group delays are similarly calculated for different state-of-the-art filters that can be used in the FOFDM based LDACS, and the comparative analysis is presented in the Table~\ref{com}. 
\begin{table}[h!]
\centering
\captionsetup{justification=centering}
		\caption{Complexity and Group Delay Comparison}
		\vspace{-0.2cm}
	\renewcommand{\arraystretch}{1}
\begin{tabular}{|m{2.5cm}|m{1.5cm}|m{1.5cm}|m{1.5cm}|}
\hline
\multirow{2}{*}{\textbf{Filters based on}}                                                                & \multirow{2}{*}{\textbf{\begin{tabular}[c]{@{}c@{}}Number of \\ Multipliers\end{tabular}}} & \multicolumn{2}{c|}{\textbf{Group Delay}} \\ \cline{3-4} 
&  & \textbf{In samples}      & \textbf{In $\mu$s }    \\ \hline
PM algorithm \cite{13, arxn}                                                          & 101                                                                               & 100             & 25             \\ \hline
LS technique \cite{13}                                                          & 75                                                                                & 74              & 18.5           \\ \hline
\multicolumn{1}{|l|}{Traditional IFIR \cite{14}} & 46                                                                                & 110.5           & 27.625         \\ \hline
\multicolumn{1}{|l|}{Generalized IFIR \cite{15} }  & 38                                                                                & 131             & 32.75          \\ \hline
\multicolumn{1}{|l|}{\begin{tabular}[c]{@{}c@{}}Proposed Filter\end{tabular}}                          & 25                                                                                & 85              & 21.25          \\ \hline
\end{tabular}
\label{com}
\end{table}
\vspace{-0.2cm}

We can observe from Table~\ref{com}, that the proposed filter offers 75.25\%, 45.65\%, 34.21\% and 66.67\% reductions in multiplication complexity and  15\%, 23.08\%, 35.11\% lower and 14.87\% higher group delay when compared to the filters designed using the PM algorithm, traditional IFIR technique, generalized IFIR technique, and LS technique respectively. 

A detailed comparison analysis of end-to-end LDACS transceivers in terms of PSD, BER, and hardware resource utilization is presented in the next section.
    \vspace{-0.17cm}

\section{Performance Analysis on ZSoC Testbed}
For performance analysis, the proposed filter is integrated with our end-to-end LDACS transceiver realized on the Xilinx Zynq ZSoC ZC706 platform, and the sample rate is set to 4 MHz. This is accomplished using MATLAB HDL Coder and Verifier Toolboxes. Please refer to \cite{arxn} for additional implementation details. Next, the transceiver is integrated with RF front-end for performance analysis in a real radio environment. The RF front-end is designed using building blocks of the MATLAB RF Toolbox. At the transmitter, RF front-end consists of digital up-converter, analog filtering, power amplifier, followed by the RF transmission. The transmission frequency can be set to anywhere in the $L$-band (960-1164 MHz) and it is set to 985 MHZ for the results discussed here. At the receiver, we need low-noise amplifier, analog filtering, and digital down-converter to get the desired baseband signal. Since RF front-end introduces phase noise, additional pilot reference signal based phase correction is added. The output of the phase correction block is passed to the proposed filter followed by baseband signal processing, as shown in Fig.~\ref{blck}.

\subsection{Power Spectral Density (PSD) Comparison}
\begin{figure}[!h]
\vspace{-0.2cm}
    \centering
    \subfloat[]{\includegraphics[scale=0.27]{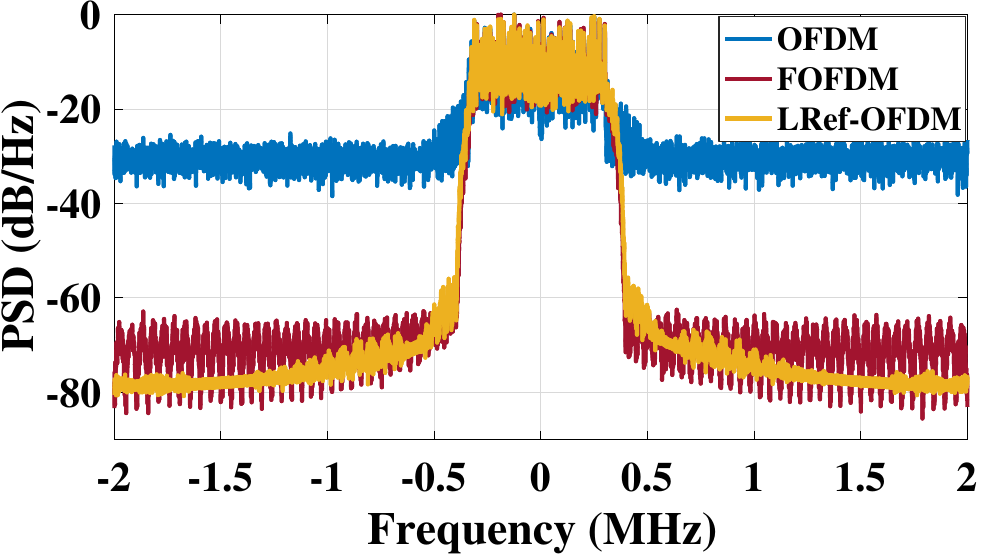}%
        \label{psd_1}}
    \subfloat[]{\includegraphics[scale=0.27]{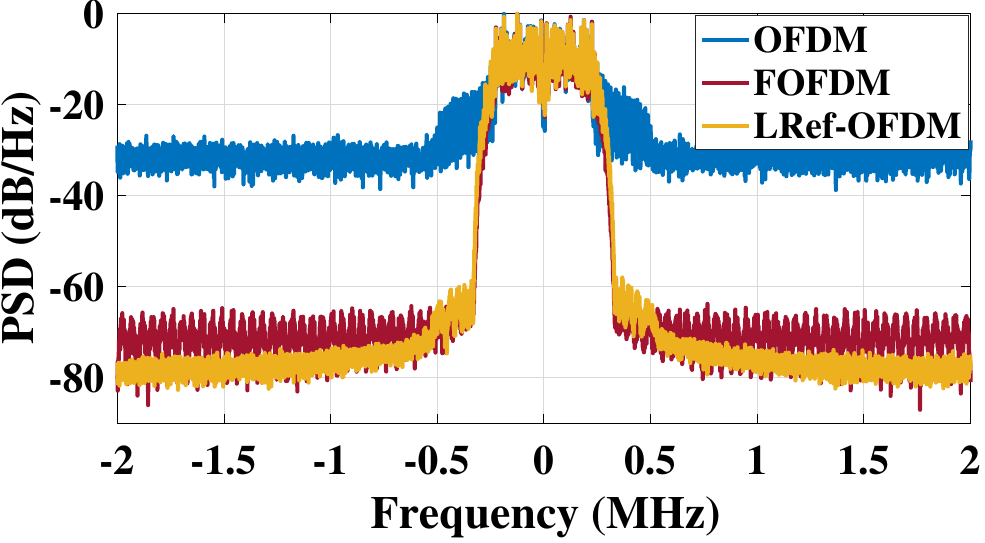}%
        \label{beb}}
        \vspace{-0.2cm}
        \caption{The PSD comparison of various waveforms for two different signal BWs, (a) 732 kHz, and (b) 498 kHz and three different channels.} 
        
        \label{psdb}
        \vspace{-0.1cm}
\end{figure}

\vspace{-0.1cm}
In Fig.~\ref{psdb}, we present the PSD comparison of OFDM, FOFDM, and LRef-OFDM based LDACS transceivers implemented using 16 bit WL. The LRef-OFDM and FOFDM based LDACS offer higher OOBA than standardized OFDM based LDACS. Higher OOBA leads to lower interference to legacy DME signals as well as enables wider transmission BW of up to 732 kHz compared to only 498 kHz in existing LDACS. This results in around 50\% improvement in the spectrum utilization. 
\vspace{-0.2cm}

\begin{figure}[!h]
    \centering
    \vspace{-0.2cm}
    \subfloat[]{\includegraphics[scale=0.27]{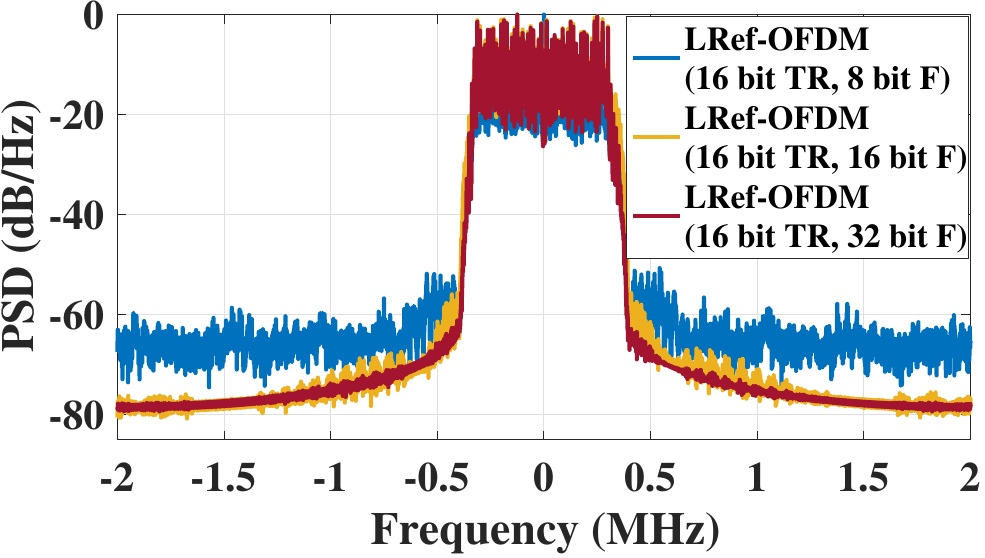}%
        \label{fa}}
    \subfloat[]{\includegraphics[scale=0.27]{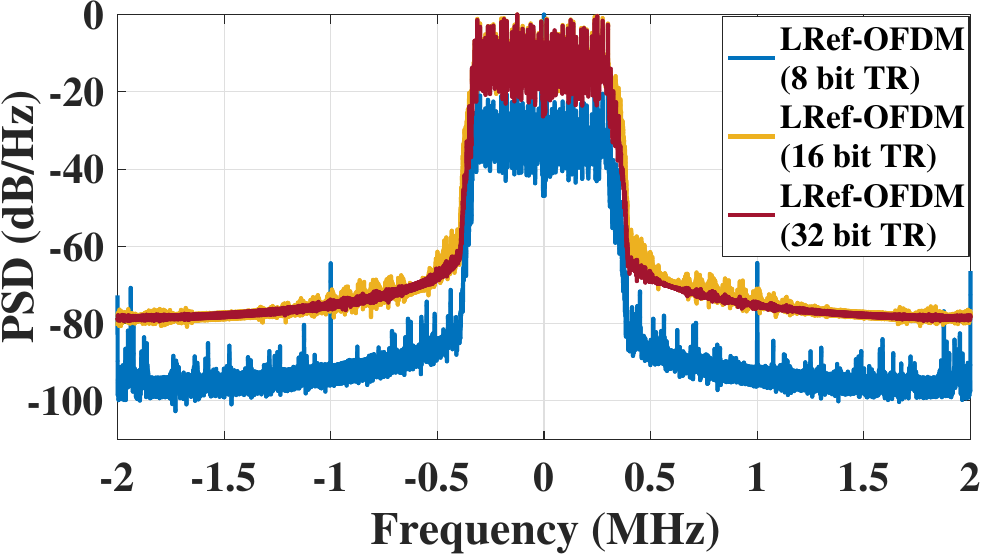}%
        \label{fb}}
        \vspace{-0.2cm}
        \caption{PSD comparison of different WLs in LRef-OFDM based LDACS. Note: TR and F in the legends refer to transceiver and filter respectively.}
            \label{fc}
            \vspace{-0.3cm}
\end{figure}

Next, we study the effect of WL on the PSD. To understand the impact of varying WL settings for the filter as well as the complete transceiver, we consider two scenarios: (1)~Proposed filter with WL of \{8,16,32\} bits and rest of the transceiver blocks with WL of 16 bits, and (2) Complete transceiver with the WL of \{8,16,32\} bits. Due to space constraints and to avoid repetitive results, we consider only LRef-OFDM based LDACS in Fig.~\ref{fa} and Fig.~\ref{fb}. In both scenarios, it can be observed that the PSD is almost identical for WL of 16 and 32 bits. However, the PSD  degrades substantially when the WL of the entire transceiver is reduced to 8 bits, as shown in Fig.~\ref{fb}. The interpretation of these results can be stated as the LRef-OFDM based LDACS system can even be implemented with lower filter WL to meet the application-specific complexity constraints. Similar results are also observed for FOFDM based LDACS, while the OFDM based LDACS needs minimum WL of 16 bits for a complete transceiver.

\vspace{-0.2cm}
\subsection{Bit Error Rate (BER) Comparison}
\begin{figure}[!h]
    \centering
    \vspace{-0.2cm}
    \subfloat[]{\includegraphics[scale=0.43]{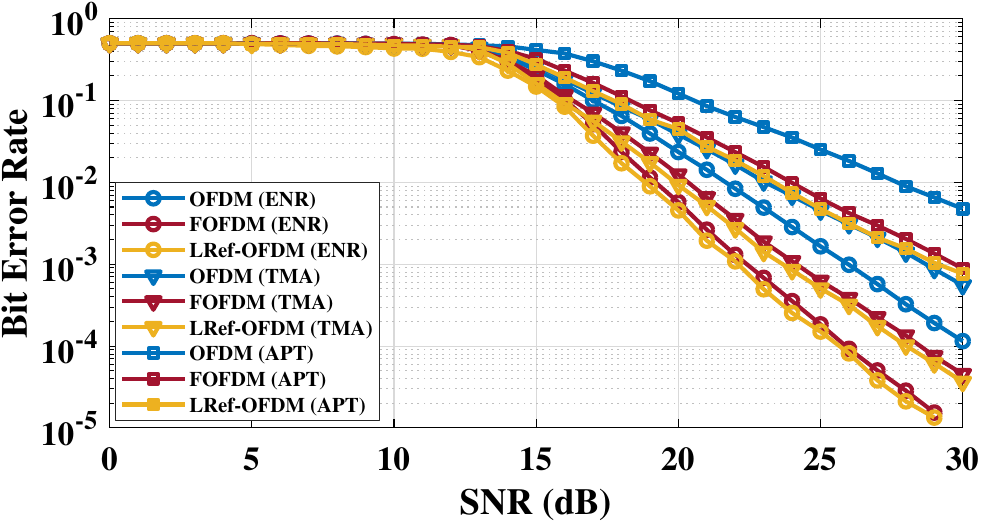}%
        \label{be7}}
\\\vspace{-0.35cm}
    \subfloat[]{\includegraphics[scale=0.43]{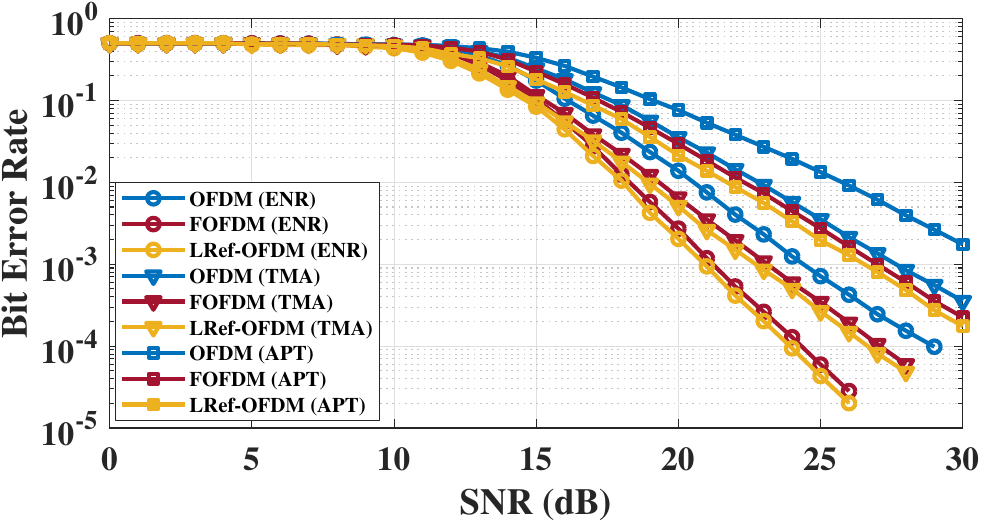}%
        \label{be4}}
        \vspace{-0.2cm}
        \caption{The BER comparison of transceivers for three different LDACS channels and two different signal BWs, (a) 732 kHz, and (b) 498 kHz.}
            \label{ber}
            \vspace{-0.2cm}
\end{figure}

\vspace{-0.1cm}
Next, we analyze the BER performance in the presence of three LDACS specific wireless channels: enrouting (ENR), airport/taxi (APT), and terminal maneuvering area (TMA). Compared to simulation-based BER discussed in literature \cite{Nr}, our analysis considers the effect of interference from legacy DME signals, impairments due to RF front-end, and different WLs. The BER analysis is done for two different signals BWs, 732 kHz and 498 kHz, and corresponding plots are shown in Fig.~\ref{be7} and Fig.~\ref{be4} respectively. It can be observed that LRef-OFDM based LDACS does not have any significant degradation in BER when compared with FOFDM based LDACS as both employ filtering to improve OOBA and to reduce the interference from DME. The errors due to RF front-end are mitigated via a phase correction block. As expected, BER of OFDM based LDACS suffers due to severe interference from DME, and its transmission BW has thus been limited to 498 kHz by the standardization committee.

\begin{figure}[!h]
\vspace{-0.2cm}
    \centering
    \captionsetup{justification=centering}
    \includegraphics[scale=0.43]{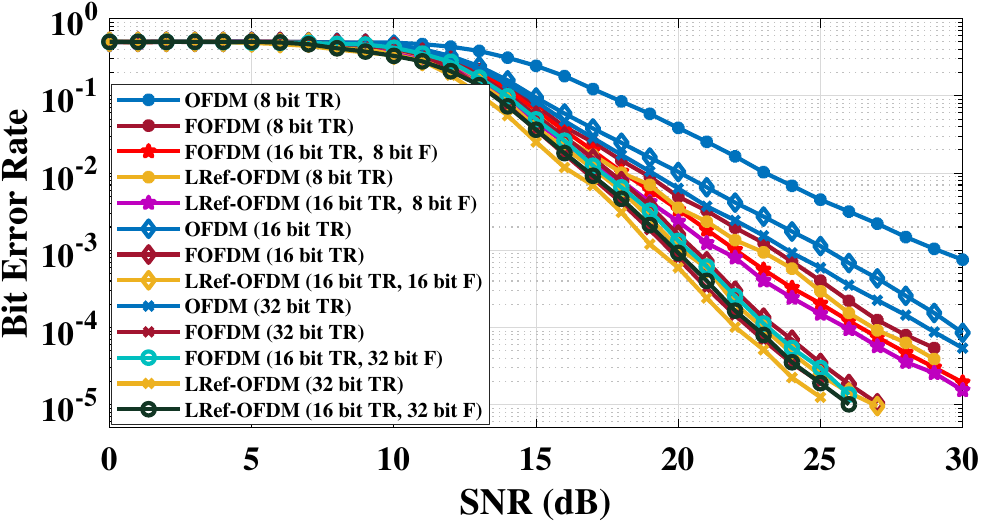}
    \vspace{-0.3cm}
    \caption{The BER comparison of various LDACS transceiver implementations for different fixed-point word lengths}
    \label{fi}
    \vspace{-0.2cm}
\end{figure}

In Fig.~\ref{fi}, we study the effect of WL on the BER performance of the LDACS transceiver. Similar to the PSD analysis, we consider two scenarios. With the decrease in WL of the filter and transceiver, the BER degrades. It can be observed that the WL of 8-bit may not be a good choice for the transceiver. However, we can have 16-bit transceiver with 8-bit filter, which also offers acceptable PSD performance. A similar analysis can be performed for each transceiver block, thereby reducing the complexity significantly without compromising on the BER and PSD. Thus, experimental BER analysis on ZSoC offers insights on the performance in real radio environment, which is otherwise not possible in simulation-based analysis. 


The LRef-OFDM based LDACS thus offers better PSD and BER performance than standardized OFDM based LDACS. It provides higher spectrum utilization due to wider transmission BW, and support for various BWs can enable LDACS to offer multiple services ranging from text, audio, to multimedia. When compared to FOFDM based LDACS, proposed LRef-OFDM based LDACS offers identical performance. Next, we compare the complexity of the LDACS transceivers.




\subsection{Resource Utilization and Power Consumption Comparison}


In this subsection, we compare the resource utilization and power consumption of the proposed filter and the PM algorithm based filter implemented on the ZC706. We also compare the same for one variant of the OFDM, FOFDM, and LRef-OFDM based LDACS transceiver corresponding to 128 point FFT. For detailed comparison of all the nine transceiver configurations one can refer to \cite{arxh}. The resource utilization comparison is presented in the Table~\ref{hd}. We can observe from the Table~\ref{hd} that the proposed filter utilizes fewer resources than the PM based filter of WL 16  bit, and hence the proposed LRef-OFDM based LDACS transceiver has lower area requirement than FOFDM based LDACS transceiver. Also, as expected, the OFDM based LDACS transceiver shows the least resource utilization due to lack of an extra filtering module. 

\begin{table}[h!]
\centering
		\captionsetup{justification=centering}
 		\caption{Resource Utilization Comparison on ZC706}
\begin{tabular}{|c|c|c|c|c|}
\hline
\multirow{2}{*}{\begin{tabular}[c]{@{}c@{}}LDACS Transceiver \\ Type\end{tabular}} & \multicolumn{4}{c|}{Resources utilized}                                                                                                                                                                                                                       \\ \cline{2-5} 
                                                                                      & \begin{tabular}[c]{@{}c@{}}No. of \\ Flip - Flops\end{tabular} & \begin{tabular}[c]{@{}c@{}}No. of \\ DSP48\end{tabular}   & \begin{tabular}[c]{@{}c@{}}No. of LUT \\ as memory\end{tabular} & \begin{tabular}[c]{@{}c@{}}No. of LUT \\ as logic\end{tabular} \\ \hline
PM filter                                                                             & \begin{tabular}[c]{@{}c@{}}10100 \\ (2.31 \%)\end{tabular}     & \begin{tabular}[c]{@{}c@{}}296 \\ (32.89 \%)\end{tabular} & \begin{tabular}[c]{@{}c@{}}64 \\ (0.09 \%)\end{tabular}         & \begin{tabular}[c]{@{}c@{}}5350 \\ (2.45 \%)\end{tabular}      \\ \hline
Proposed filter                                                                       & \begin{tabular}[c]{@{}c@{}}8921 \\ (2.04 \%)\end{tabular}      & \begin{tabular}[c]{@{}c@{}}181 \\ (20.11 \%)\end{tabular} & \begin{tabular}[c]{@{}c@{}}50 \\ (0.073 \%)\end{tabular}        & \begin{tabular}[c]{@{}c@{}}4114 \\ (1.89 \%)\end{tabular}      \\ \hline
\begin{tabular}[c]{@{}c@{}}OFDM based\\ LDACS transceiver\end{tabular}                & \begin{tabular}[c]{@{}c@{}}39501 \\ (9.03 \%)\end{tabular}     & \begin{tabular}[c]{@{}c@{}}570\\ (63.33 \%)\end{tabular}  & \begin{tabular}[c]{@{}c@{}}994 \\ (1.35 \%)\end{tabular}        & \begin{tabular}[c]{@{}c@{}}37541 \\ (17.23 \%)\end{tabular}    \\ \hline
\begin{tabular}[c]{@{}c@{}}FOFDM based \\ LDACS transceiver\end{tabular}              & \begin{tabular}[c]{@{}c@{}}47653\\ (10.89 \%)\end{tabular}     & \begin{tabular}[c]{@{}c@{}}812\\ (90.22 \%)\end{tabular}  & \begin{tabular}[c]{@{}c@{}}1021\\ (1.44 \%)\end{tabular}        & \begin{tabular}[c]{@{}c@{}}43439\\ (19.93 \%)\end{tabular}     \\ \hline
\begin{tabular}[c]{@{}c@{}}LRef-OFDM based \\ LDACS transceiver\end{tabular}            & \begin{tabular}[c]{@{}c@{}}46394\\ (9.82 \%)\end{tabular}     & \begin{tabular}[c]{@{}c@{}}697\\ (77.44 \%)\end{tabular}  & \begin{tabular}[c]{@{}c@{}}1009\\ (1.42 \%)\end{tabular}        & \begin{tabular}[c]{@{}c@{}}41552\\ (19.07 \%)\end{tabular}     \\ \hline
\end{tabular}
\label{hd}
\end{table}


We also compare the dynamic power and critical path delay for all the transceivers. While LRef-OFDM based LDACS (0.437 W) consumes 30.43\% more power than OFDM based LDACS (0.304 W), it consumes 14.14 \% less power than FOFDM based LDACS (0.509 W) due to proposed filter design. The critical path delay for LRef-OFDM based LDACS (12.1 ns) is lower compared to FOFDM based LDACS (12.5 ns), and this can be advantageous (higher throughput) if multi-clock designs are implemented. Additionally, we compare the complexity for different WLs of OFDM and LRef-OFDM based LDACS transceiver in Table.~\ref{hd1}. We observed that resource utilization and power consumption increase with the increase in the WL. To summarize, LRef-OFDM based LDACS offers better OOBA and BER performance than OFDM based LDACS, along with lower implementation complexity and power consumption than FOFDM based LDACS. This makes it an attractive alternative for future air-ground communications. 



\begin{table}[h!]
\centering
 \vspace{-0.1cm}
		\captionsetup{justification=centering}
 		\caption{Resource Utilization Comparison for Transceiver's different Word Lengths on ZSoC ZC706}
\begin{tabular}{|c|c|c|c|c|}
\hline
\multirow{2}{*}{\begin{tabular}[c]{@{}c@{}}Resources\\ Utilized\end{tabular}} & \multicolumn{2}{c|}{OFDM LDACS}                            & \multicolumn{2}{c|}{LRef-OFDM LDACS}                         \\ \cline{2-5} 
                                                                              & 8-bit WL                                                   & 32-bit WL                                                  & 8-bit WL                                                   & 32-bit WL                                                  \\ \hline
\begin{tabular}[c]{@{}c@{}}No. of \\ Flip-Flops\end{tabular}                  & \begin{tabular}[c]{@{}c@{}}33514\\ (7.66 \%)\end{tabular}  & \begin{tabular}[c]{@{}c@{}}51254\\ (11.72 \%)\end{tabular} & \begin{tabular}[c]{@{}c@{}}38421\\ (8.79 \%)\end{tabular}  & \begin{tabular}[c]{@{}c@{}}58624\\ (13.40 \%)\end{tabular} \\ \hline
\begin{tabular}[c]{@{}c@{}}No. of LUT \\ as logic\end{tabular}                & \begin{tabular}[c]{@{}c@{}}31421\\ (14.42 \%)\end{tabular} & \begin{tabular}[c]{@{}c@{}}47264\\ (21.69 \%)\end{tabular} & \begin{tabular}[c]{@{}c@{}}36283\\ (16.65 \%)\end{tabular} & \begin{tabular}[c]{@{}c@{}}53234\\ (24.43 \%)\end{tabular} \\ \hline
\end{tabular}
\label{hd1}
 \vspace{-0.1cm}
\end{table}
%
    \vspace{-0.17cm}
\section{Conclusion}
In this paper, we presented LRef-OFDM based LDACS using low complexity reconfigurable filter design approach. With extensive experimental results on a hardware testbed, we validated its superiority over OFDM based LDACS in terms of PSD and BER performance along with tunable bandwidth. It also offers a lower area and power complexity than FOFDM based LDACS. Our work considered in-depth experimental analysis via different word lengths, RF impairments, and LDACS channels compared to existing simulation-based studies. Future work includes hardware implementation of pulse blanking technique to mitigate DME interference, and investigating solutions to improve the BER performance of the proposed LRef-OFDM based LDACS.


\onecolumn
\large
\section*{\huge{Appendix}}

In this appendix, firstly we will give an idea about some recent state of the art filters mainly based on interpolation method. Then we provide the design details for the proposed low complexity reconfigurable filter followed by the mathematical analysis for the proposed LRef-OFDM based LDACS transceiver. In the end we also present some additional resource and power utilization results for 
various transceiver configurations based on PS/PL boundary.

\section{Literature Review}
Numerous filter design approaches employing the interpolation technique have been proposed to obtain low complexity FIR filters. For example, a computationally efficient digital filter based on interpolation and frequency-response-masking (FRM) technique is presented in \cite{Wu}. In this paper, the designed filter is optimized in minimax sense by jointly optimizing the subfilters involved using a convex-concave procedure (CCP). Authors also extend the work to the design of FRM filters that simultaneously promotes sparsity of the filter coefficients to reduce implementation complexity. 

Similarly, interpolation based narrow-band sparse FIR filters and centrosymmetric bandpass filters are designed in \cite{WC}. The design method is realized by cascading a model filter with a sparse masking filter. In this technique, the model filter is first designed and then interpolated to generate the desired impulse response replica. A sparse masking filter is used to mask the extra unwanted passbands. 

A low complexity 17-band non-uniform IFIR filter bank for digital hearing aid applications is designed in \cite{TD}. In this filter, different levels of interpolations are done on these filters, to create various bands.

The filtered OFDM can be applied in several other communication environments suitable for next generation wireless communication system. Such as, to achieve better spectrum leakage performance in 5G wireless communication, authors in \cite{LY} have proposed a Nuttall’s Blackman-Harris windowed F-OFDM system, and analyzes the performance of different window functions in the F-OFDM. They have also compared the proposed filter’s performance with the hamming windowed filter and claims to achieve better PSD and BER performance with the proposed filter.

Similarly, a winnowed sync filter for filtered OFDM to reduce the out-of-band emission is discussed in \cite{AT}. Authors have proposed an optimal filter choice by analysing the system performance by employing six windowing functions such as: Hann, Hamming, Kaiser, Barlett, Blackman, Flat-top.  They further study the performance of a two-user uplink scenario where the two uplink users pass the respective OFDM modulated data through time domain filters of appropriate bandwidths. According to the author’s analysis Kaiser Window gives the best performance in terms of BER and OOB reduction.

Further, a novel resource block (RB) filtered OFDM (RBF-OFDM) is proposed in \cite{JL}, that segregates the entire available spectrum into different blocks (RBs) and filters individual signal transmitted on each block. The performance of RB F-OFDM is similar to filtered OFDM with the additional support for non-contiguous spectrum under channels with moderate delay spread and adjacent channel interference (ACI). 

An interpolated band-pass method (IBM) based narrow-band finite impulse response (FIR) filter for 5G cellular network is proposed in \cite{SR}. The proposed filter consists of different sub-filters such as prototype filter, $H_{a}(e^{j\omega})$, complementary prototype filter $H_{c}(e^{j\omega})$. The approach considers a band-pass filter (BPF), $H_{bp}(e^{j\omega})$ placed in between prototype and complementary prototype filter pair.  Authors in this paper claims that the proposed IBM based narrow-band filter accomplishes shorter out-of-band emission (OOBE) without affecting the BER performance while compared to the other narrow-band filters.

Authors in paper \cite{YQ}, proposed filters constructed by filter banks to transmit and receive F-OFDM signals to adapt the flexible numerology of 5G communication. The proposed filter bank supports multiple sub-bands and consists an analysis filter bank (AFB) and a synthesis filter bank (SFB). Each filter bank is composed of a set of modulated filters. Further the sub-band filters are implemented by using the poly-phase structure to reduce the computational complexity.

Next, we present the design details of the proposed low complexity reconfigurable filter.

\section{Low Complexity Reconfigurable Filter Design}
The proposed filter is a interpolation based multi-stage filter. The type and parameters of every stage of the proposed filters is designed to shape the  transmission  spectrum  so as it  meets  the  stringent  LDACS  spectral  mask, BER and PSD requirements.  At the same time, overall filter complexity (in terms of total number of unique multipliers) and group delay should be as small as possible.

In an interpolation based multi-stage filter design approach like the one followed in our work, three parameters are interlinked – (1) number of cascaded filter stages, (2) interpolation factors for every filter stage, and (3) the prototype filter orders for every filter stage. The overall filter complexity (in terms of total number of unique multipliers) of a multi-stage filter is dependent on the number of stages and interpolation factor for each stage. The interdependence between these parameters can be described as – 
\begin{itemize}
    \item If number of cascaded filter stages is kept fixed, then choosing a higher interpolation factor for the first stage filter (thus lower order and lesser multipliers for corresponding prototype filter) results in the requirement of higher order masking filters for the subsequent stages (due to more stringent masking requirements to remove the unwanted subbands that are more in number and closer together). Conversely, choosing a lower interpolation factor for the first stage filter can help to lower the filter order for the subsequent stages’ masking filters (due to relatively relaxed masking requirements to remove the unwanted subbands that are lesser in number and farther away), but this choice requires a relatively higher filter order and thus more multipliers for the prototype filter in the first stage itself. The overall filter complexity in terms of the total required number of multipliers thus needs to be minimized by considering the full set of interpolation factors in all the stages appropriately. 
    \item In the LDACS scenario, the spectrum mask mandates varying attenuation levels for different frequencies wherein highest attenuation levels are required furthest away from the LDACS signal. In our design approach, we use three cascaded filter stages so that we can achieve the different attenuation levels using aggregation of attenuations provided by the individual filters. This helps to keep the filter orders (and thus required number of multipliers) low in every stage while ensuring that the overall filter frequency response satisfies the LDACS spectral mask requirements. 
    \item Meanwhile, when deciding the interpolation factors, it is to be noted that the maximum interpolation factor $I_{max}$ that can be chosen for a filter stage with final desired normalized stopband edge frequency Fstop of the desired subband in the interpolated frequency response is given by $I_{max} =  \left\lfloor \frac{1}{F_{stop}} \right\rfloor$. This upper limit ensures that the prototype filter design corresponding to the chosen interpolation factor for that stage is feasible, i.e., stopband edge frequency specification of the protype filter is less than 1. In our filter design, the maximum possible interpolation factor for the first stage filter is obtained as 4 by considering the frequency edge specifications corresponding to the widest supported LDACS signal bandwidth, i.e., 732 kHz.  
\end{itemize}

By considering the above points in our work, we concluded to design a three stage filter with interpolation factors {4, 2, 1} having the filter order as 26, 26, 14 provides the lowest overall filter complexity.

To explain further using a specific example, table provides a comparison of few of the possible multi-stage filter designs for our variable bandwidth LDACS system. It can be noted how the lowest overall filter complexity in terms of total number of multipliers required is achieved by Option 4 which is our final choice of design discussed in the paper. Note that the LDACS spectrum mask mentioned in LDACS standard provides the desired attenuation levels for various values of frequency difference with respect to carrier frequency. Since the spectrum mask only provides numerical values without any mathematical model, interpolation factors, number of stages and filter order are selected empirically to meet the desired attenuation constraints and minimize the computational complexity.

\begin{table}[h!]
\resizebox{\textwidth}{!}{%
\begin{tabular}{|c|c|c|c|c|c|c|}
\hline
\textbf{Option}                                              & \textbf{\begin{tabular}[c]{@{}c@{}}Number of \\ stages\end{tabular}} & \textbf{\begin{tabular}[c]{@{}c@{}}Interpolation \\ factor for \\ each stage\end{tabular}} & \textbf{\begin{tabular}[c]{@{}c@{}}Filter\\  order for \\ every stage\end{tabular}} & \textbf{\begin{tabular}[c]{@{}c@{}}Type of \\ filters in\\ every stage\end{tabular}}    & \textbf{\begin{tabular}[c]{@{}c@{}}Overall filter \\ order\end{tabular}}       & \textbf{\begin{tabular}[c]{@{}c@{}}Total number of \\ unique multipliers \\ required for\\  implementation\end{tabular}} \\ \hline
1                                                            & 2                                                                    & \{4, 1\}                                                                                   & \{26, 50\}                                                                          & \{FIR, FIR\}                                                                            & (26x4) + 50 = 154                                                              & 14+26 = 40                                                                                                               \\ \hline
2                                                            & 2                                                                    & \{3, 1\}                                                                                   & \{44, 24\}                                                                          & \{FIR, FIR\}                                                                            & (44x3) + 24 = 156                                                              & 23+13 = 36                                                                                                               \\ \hline
3                                                            & 2                                                                    & \{2, 1\}                                                                                   & \{104, 14\}                                                                         & \{FIR, FIR halfband\}                                                                   & (104x2) + 14 = 222                                                             & 53+4 = 57                                                                                                                \\ \hline
\textbf{\begin{tabular}[c]{@{}c@{}}4\\ (Final)\end{tabular}} & \textbf{3}                                                           & \textbf{\{4, 2, 1\}}                                                                       & \textbf{\{26, 26, 14\}}                                                             & \textbf{\begin{tabular}[c]{@{}c@{}}\{FIR, FIR halfband,\\  FIR halfband\}\end{tabular}} & \textbf{\begin{tabular}[c]{@{}c@{}}(26x4) + (26x2) \\ + 14 = 170\end{tabular}} & \textbf{14+7+4 = 25}                                                                                                     \\ \hline
\end{tabular}}
\end{table}

The interpolated frequency response can be given by:
\begin{equation}
    H_{Ip}(z)=\sum_{n=0}^{\frac{N}{2}-1}h_n[z^{-Mn}+z^{-M(N-n)}]+h_{\frac{N}{2}}z^{\frac{-Mn}{2}} 
    \label{1}
\end{equation}

where, $M$ is the interpolation factor, $N$ is the filter order, $h_{0}, h_{1}$ \ldots $h_{\frac{N}{2}}$ are the unique filter coefficients of an $N^{th}$ order FIR filter.

\begin{figure}[!h]
  \vspace{-0.3cm}
    \centering
    \captionsetup{justification=centering}
    \includegraphics[scale=0.6]{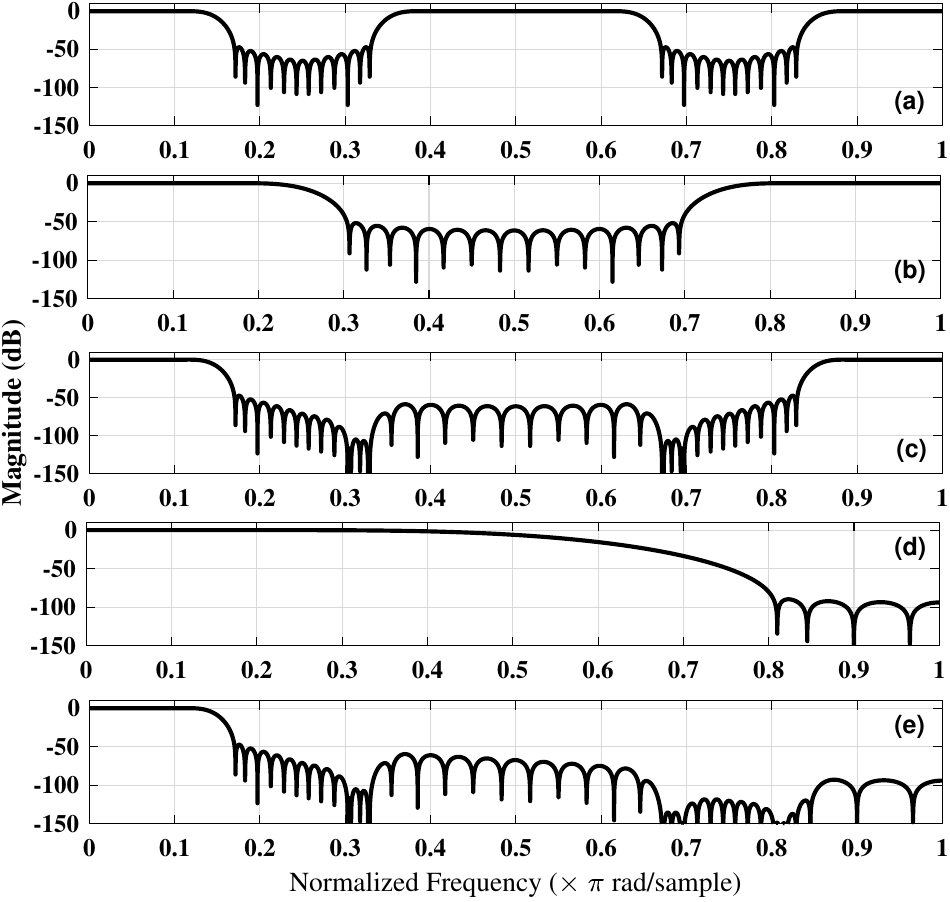}
     \vspace{-0.2cm}
    \caption{Frequency response of (a) first stage filter (b) second-stage filter (c) output of second stage filter ($H_I(z).H_{II}(z)$) (d) third stage filter (e) Output of third stage filter ($H_I(z).H_{II}(z).H_{III}(z)$)}
    \label{resp31}
\end{figure}

The lowpass prototype filter I (stage 1) are interpolated by a factor 4 and has the passband edge ($Fp_{1}$) as four times that of the passband edge of the signal ($Fp_{s}$) to be filtered $Fp_{1}=4*Fp_{s}$. Similarly, the stopband edge ($Fs_{1}$) of Filter I is selected as $4*Fs_{s}$ (stopband edge of the signal to be filtered). The interpolated multi-band frequency response is shown in Fig.~\ref{resp31} (a) and can be given by:

\begin{equation}
    H_{I}(z)=\sum_{n=0}^{12}h_n[z^{-4n}+z^{-4(26-n)}]+h_{13}z^{-2n} 
\label{de}
\end{equation}

The Filter II (stage II) and III (stage III) are used as masking filters to remove the unwanted subbands and are designed by employing interpolation factor 2 and 1 (interpolation factor 1 signifies no interpolation operation) respectively. Based on the design of Filter I, a reference frequency ($F_{m}$) is computed and the passband and stopband edge frequencies of Filter II and Filter III are chosen based on $F_{m}$. The reference frequency $F_{m}$ is selected based on the supported transmission BWs and is chosen to be the stopband edge frequency of Filter I corresponding to the widest supported transmission BW. Once this reference frequency is selected, we define the passband edge frequency for Filter II and Filter III as $Fp_2=\frac{F_m}{2}$ and $Fp_{3}=\frac{F_{m}}{4}$ respectively. Furthermore, to obtain both these masking filters as halfband filters, the stopband edge frequency is chosen as $Fs_{2}=1-Fp_{2}$ and $Fs_{3}=1-Fp_{3}$ for Filter II and Filter III respectively and the frequency responses ($H_{II}(z), H_{III}(z)$) are illustrated in Fig.~\ref{resp31} (b) and (d) respectively and can be represented as :
\begin{equation}
    H_{II}(z)=\sum_{n=0}^{12}h_n[z^{-2n}+z^{-2(26-n)}]+h_{13}z^{-n} 
\label{e}
\end{equation}
    
\begin{equation}
    H_{III}(z)=\sum_{n=0}^{6}h_n[z^{-n}+z^{-(14-n)}]+h_{7}z^{\frac{-n}{2}}
\label{f}
\end{equation}

The output of the stage II filter $H_I(z).H_{II}(z)$ is shown in Fig.~\ref{resp31} (c). 

Furthermore, the masking filter stage of the proposed approach are fixed and independent of the transmission bandwidth. This is done by considering the worst-case passband edge and stopband edge specifications of the masking filters (Filter II and III). The text in Section II on page 2 explaining the sub-filter design has been updated in the reviewed manuscript accordingly.

\begin{figure}[!h]
  \vspace{-0.3cm}
    \centering
    \captionsetup{justification=centering}
    \includegraphics[scale=0.6]{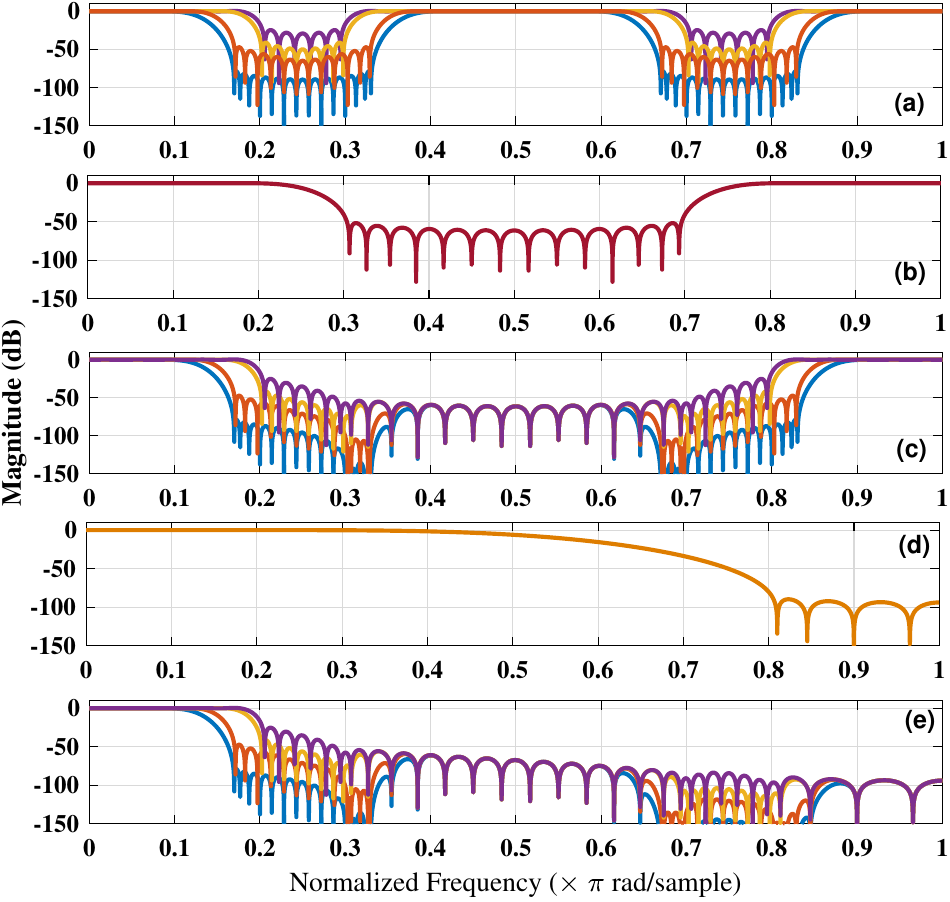}
     \vspace{-0.2cm}
    \caption{Variable BW frequency response of (a) first stage filter (b) second-stage filter (c) cascaded output of second stage filter ($H_I(z).H_{II}(z)$) (d) third stage filter (e) cascaded output of third stage filter ($H_I(z).H_{II}(z).H_{III}(z)$)}
    \label{resp41}
\end{figure}

To support above explanation, we have included Fig.~\ref{resp41} which shows the frequency response for the different transmission bandwidths : 342 kHz, 498 kHz, 654 kHz, 732 kHz. The multi-band frequency response for filter I (interpolated by 4) and filter II (interpolated by 2) is shown in Fig.~\ref{resp41} (a) and (b) respectively. Fig.~\ref{resp41} (c) shows the output obtained at the stage II by cascading the Filter I ($H_I(z)$) and Filter II ($H_{II}(z)$). The output of the stage II filter is then cascaded with the stage III filter (as shown in Fig.~\ref{resp41} (d)) to obtain the resultant filter as shown in Fig.~\ref{resp41} (e).

Next, we present the mathematical model for the proposed LRef-OFDM based LDACS transceiver.
         

\section{Mathematical illustration for LRef-OFDM based LDACS transceiver}
\subsection{LRef-OFDM based LDACS Transmitter}
A brief analysis for the transmitter is given as follows:
The discrete time domain OFDM signal corresponding to the kth subcarrier can be given as
\begin{equation}   
x[n]= \frac{1}{K} \sum \limits_{k=0}^{K-1} X_k e^{\frac{j 2\pi kn}{K}}
    \label{3}
\end{equation}
where, $K$ is the IFFT size, $n$ is the discrete time index and $X_k$ is frequency domain response of the transmitted signal at the $k^{th}$ sub-carrier. It is given by
\begin{equation}
    X_k= \sum \limits_{n=0}^{K-1} x[n] e^{\frac{-j 2\pi kn}{K}}
    \label{4}
\end{equation}
The transmitted signal $x'[n]$ is the convolution ($\ast$) of $x[n]$ and the filter $h[n]$ and can be expressed as,
\begin{equation}
    x'[n]= h[n] \ast x[n]
    \label{5}
\end{equation}

The transmitted signal then passes through the LDACS wireless channel with impulse response $c^{L}[n]$, and a DME interference signal passes through the LDACS wireless channel with impulse response $c^{D}[n]$.

\subsection{LRef-OFDM based LDACS Receiver}
The signal received at the input of the receiver r[n] can be expressed as,
\begin{equation}
    r[n]= c^L[n] \ast x'[n] + c^D[n] \ast s[n]+ \Tilde{n_0}[n]
    \label{8}
\end{equation}

where, $x'[n]$ is the transmitted signal, $s[n]$ is the DME signal, and $\Tilde{n_0}[n]$ is the zero-mean additive white Gaussian noise.

In LDACS environment, both channels $c^L[n]$ and $c^D[n]$ assumed to have identical statistics such that
\begin{equation}
    c^L[n]=\sum_{l=1}^{L} c^L_l \delta[n-l] ~and~  c^D[n]=\sum_{l=1}^{L} c^D_{l} \delta[n-l]
    \label{9}
\end{equation}
Where $L$ is the total number of channel taps, $c^L_l$ and $c^D_{l}$ are the impulse responses of the channel faced by LDACS and DME signal of the $l^{th}$ path, respectively.  The channels are assumed to be time-invariant for each transmitted OFDM symbol.

\begin{equation}
r'[n]= h[n] \ast c^L[n] \ast h[n] \ast x[n] + c[n] \ast c^D[n] \ast s[n]+ c[n] \ast \Tilde{n_0}[n]    
 \label{10}
\end{equation}

After doing FFT, the received signal $r'[n]$ is converted into the frequency domain and can be represented as $R'_k$ at $k^{th}$ subcarrier:
\begin{equation}
    R'_k= H_k C_k H_k X_k + H_k C_{d_k} S_k + H_k \Tilde{N_0}{k}
    \label{11}
\end{equation}

where, $C_k$ , $C_{d_k}$ are the LDACS and DME channel frequency responses at the $k^{th}$ subcarrier respectively can be given as
\begin{equation}
    C_k= \sum_{l=1}^{L} c^L_{l} e^\frac{-j2\pi kl}{N} ~and~ C_{d_k}= \sum_{l=1}^{L} c^D_{l} e^\frac{-j2\pi kl}{N} 
\label{Hnew}
\end{equation}
we obtain $H_{k}$ by taking K-point FFT of the zero padded filter impulse response where
\begin{equation}
        H_k= W^H. [h[n]]
        \label{12}
\end{equation}

Here, $W$ is the $K$-point FFT matrix. 

The impulse response of the proposed filter $h[n]$ can be calculated by taking Inverse ZT of the resultant cascaded filter $H(z)$.
\begin{equation}
        h[n]= \Zstroke^{-1}[H(z)]
        \label{12}
\end{equation}

where, the proposed LRef filter response $H(z)$ can be given as,
\begin{equation}
    H(z) = H_{I}(z).H_{II}(z).H_{III}(z)
    \label{a}
\end{equation}
$H_{I}(z),H_{II}(z),H_{III}(z)$ are as given in eq.~\ref{de},~\ref{e},~\ref{f}

 We finally realize the BER performance of the $k^{th}$ received symbol for M-QAM can be expressed as,
\begin{equation}
    P_{e_{MQAM}}^{\lambda,\lambda_d}(k) \cong \frac{4}{log_2 M} \left (1-\frac{1}{\sqrt{M}}  \right ) \sum_{i=1}^{\sqrt{M}/2} Q(2i-1) \times \sqrt{\frac{3 log_2 M H_{k}^{2} \lambda_k P}{\left (M-1  \right ) \left (P_{\tilde{N_0}} + \lambda_{d_k} P_{DME} \right )}}
    \label{23}
\end{equation}
where $P$ is the transmitted signal power, $P_{DME}$ is the DME signal power. A detailed derivation for $P_{DME}$ is given in  \cite{NA}. erfc(.) is a complex error function, therefore, BER averaged across the fading channel can be expressed as,
\begin{equation}
    \nonumber P_{e_{MQAM}}(k) = E[P_{e_{MQAM}}^{\lambda,\lambda_d}(k)] \cong\int_{0}^{\infty}\int_{0}^{\infty} P_{e_{MQAM}}^{\lambda_1,\lambda_2}(k)~ \times p_{\lambda}(\lambda) d\lambda ~p_{\lambda_{d}}(\lambda_d) d\lambda_{d}
    \label{24}
\end{equation}

The average BER across all the subcarriers is given by
\begin{equation}
    P_{e_{MQAM}} = \frac{1}{K} \sum_{k=0}^{K-1} P_{e_{MQAM}}(k)  
    \label{25}
\end{equation}

Next, we present the additional results comparing various transceiver configurations corresponding to different boundary between PS and PL.

\section{Experimental Results: Resource Utilization and Power Consumption Comparison}
Here, we present the resource utilization and power consumption results for the nine different configurations implemented via hardware software co-design approach \cite{arxn}. The comparison is done for OFDM, FOFDM (PM based filtered OFDM), and the proposed LRef-OFDM based LDACS transceivers. Here, we consider 16 bit word length and the highest bandwidth possible 732 kHz for the comparison. All results are  obtained  after  realizing  the  transceiver  on  ZC706  from Xilinx. Since  V1  configuration  is  realized  completely  in  PS,  PL resource  utilization  results  are  omitted.  In  V2,  FOFDM and LRef-OFDM resource utilization is due to the filtering block realized in PL. Later, one by one block is transferred to the PL in the subsequent configurations. 

We can observe from the Table~\ref{result} that the proposed LRef-OFDM based LDACS transceiver has lower area and power requirement than FOFDM based LDACS transceiver. Also, as expected, the OFDM based LDACS transceiver shows the least resource utilization and dynamic power due to lack of an extra filtering module.

\begin{table*}[!t]
	\centering
		\captionsetup{justification=centering}
		\caption{Resource Utilization and Power Consumption of Transceiver on ZSoC}
		\vspace{-0.2cm}
	\renewcommand{\arraystretch}{1}
	\begin{tabular}{|C{1.8cm}|C{1.4cm}|C{1.2cm}|C{1.2cm}|C{1.2cm}|C{1.2cm}|C{1.2cm}|C{1.2cm}|C{1.3cm}|C{1.3cm}|} 	
		\hline
		\textbf{Parameter} & \textbf{Waveform} & \textbf{V2} & \textbf{V3} & \textbf{V4} & \textbf{V5}& \textbf{V6} & \textbf{V7} & \textbf{V8} & \textbf{V9}\\
		\hline
		\textbf{No. of Flip-Flops} & OFDM & N/A & 14851 (3.39\%) & 31891 (7.29\%) & 32385 (7.41\%) & 32981 (7.54\%) & 34351 (7.86\%) & 38127 (8.72\%) & 39501 (9.03\%)\\
		& FOFDM & 10100 (2.31\%) & 30124 (6.89\%) & 37912 (8.67\%) & 40174 (9.19\%) & 40927 (9.36\%) & 42129 (9.63\%) & 44852 (10.26\%) & 47653 (10.89\%)\\
		& LRef-OFDM & 8921 (2.04\%) &  27521 (6.29\%) & 33874 (7.75\%) & 36981 (8.46\%) & 38121 (8.72\%) & 39984 (9.15\%) & 42191 (9.65\%) & 46394 (10.61\%)\\
   
		\hline
		\textcolor{blue}{\textbf{No. of DSP48}} & OFDM & N/A & 534 (59.33\%) & 570 (63.33\%) & 570 (63.33\%) & 570 (63.33\%) & 570 (63.33\%) & 570 (63.33\%) & 570 (63.33)\\
		& FOFDM & 296 (32.89\%) & 785 (87.22\%) & 812 (90.22\%) & 812 (90.22\%) & 812 (90.22\%) & 812 (90.22\%) & 812 (90.22\%) & 812 (90.22\%) \\
		& LRef-OFDM & 181 (20.11) & 668 (74.22\%) & 697 (77.44\%) & 697 (77.44\%) & 697 (77.44\%) & 697 (77.44\%) & 697 (77.44\%) & 697 (77.44\%)\\
		\hline
		\textbf{No. of LUT as Memory}& OFDM & N/A & 396 (0.56\%) & 865 (1.23\%) & 881 (1.25\%) & 918 (1.30\%) & 922 (1.31\%) & 941 (1.34\%) & 994 (1.35\%)\\
		& FOFDM & 64 (0.09\%) & 411 (0.583\%) & 894 (1.27\%) & 913 (1.29\%) & 936 (1.32\%) & 943 (1.34\%) & 964 (1.37\%) & 1021 (1.44\%)\\
		& LRef-OFDM & 50 (0.073\%) & 392 (0.55\%) & 871 (1.23\%) & 898 (1.26\%) & 913 (1.28\%) & 931 (1.31\%) & 946 (1.33\%) & 1009 (1.42\%)\\
  
		\hline
		\textcolor{blue}{\textbf{No. of LUT as Logic}} & OFDM & N/A & 23124 (10.58\%) & 32185 (14.72\%) & 32912 (15.06 \%) & 33842 (15.48\%) & 34251 (15.67\%) & 35921 (16.43\%) & 37541 (17.23\%)\\
		& FOFDM & 5350 (2.45\%) & 26571 (12.16\%) & 33125 (15.15\%) & 34981 (16.01\%) & 36521 (16.71\%) & 37852 (17.32\%) & 41146 (18.82\%) & 43439 (19.07\%)\\
		
		& LRef-OFDM & 4114 (1.89\%) & 25124 (11.49\%) & 32742 (14.98\%) & 33521 (15.33\%) & 35215 (16.11\%) & 36259 (16.59 \%) & 40212 (18.55\%) & 41552 (19.07\%)\\

		\hline
		\textbf{No. of}& OFDM & N/A & 35 & 683 & 745 & 1144 & 1217 & 1882 & 1930\\
		\textbf{MUXes} & FOFDM & 25 & 57 & 835 & 1152 & 1401 & 1523 & 1985 & 2102\\
		& LRef-OFDM & 19 & 47 & 742 & 951 & 1261 & 1394 & 1885 & 2041 \\
		\hline
	    \textcolor{blue}{\textbf{Dynamic}} & OFDM & N/A & 0.045 & 0.285 & 0.295 & 0.297 & 0.299 & 0.301 & 0.304\\
	    \textcolor{blue}{\textbf{Power}} & FOFDM & 0.112 & 0.205 & 0.434 & 0.493 & 0.494 & 0.496 & 0.500 & 0.509\\
	    \textcolor{blue}{\textbf{in Watt}} & LRef-OFDM & 0.072 & 0.134 & 0.375 & 0.401 & 0.421 & 0.429 & 0.434 & 0.437\\	\hline
	\end{tabular}
		\vspace{-0.2cm}
	\label{result}
\end{table*}

\end{document}